\documentclass[fleqn,twoside]{article}
\usepackage{espcrc2}

\usepackage{graphicx}
\usepackage[figuresright]{rotating}

\newcommand{\AmS}{{\protect\the\textfont2
  A\kern-.1667em\lower.5ex\hbox{M}\kern-.125emS}}

\title{Numerical evaluation of general massive  2-loop self-mass master 
integrals from differential equations}

\author{M. Caffo\address[INFN]{INFN, Sezione di Bologna, \\
                               I-40126 Bologna, Italy}
 \kern-3pt\address[DFBO]{Dipartimento di Fisica, Universit\`a di Bologna, \\
                               I-40126 Bologna, Italy}
           \thanks{Speaker at ACAT'2002,
                VIII International Workshop on Advanced Computing and 
                Analysis Techniques in Physics Research,
                Moscow, 24-28 June 2002. },
        H. Czy{\.z}\address{Institute of Physics, University of Silesia, \\
        PL-40007 Katowice, Poland}
        and
        E. Remiddi\addressmark[DFBO]\addressmark[INFN] }
       
\begin{document}

\begin{abstract}
The system of 4 differential equations in the external invariant satisfied 
by the 4 master integrals of the general massive 2-loop sunrise self-mass 
diagram is solved by the Runge-Kutta method in the complex plane. 
The method offers a reliable and robust approach to the direct and precise 
numerical evaluation of Feynman graph integrals.
\vspace{1pc}
\end{abstract}
\maketitle

The relevance of the higher order calculations for the comparison with 
nowadays precision measurements in high energy physics is well known 
and comprehensively presented by G. Passarino in this conference. 

Therefore can be of some interest the exploitation of an alternative 
method (but still in the context of the integration by part identities 
and master integrals (MI) \cite{TkaChet}) to the more common direct 
integration method for the numerical evaluation of the MI.

The method uses directly the differential equations. Starting from the 
integral representation of the MI, related to a certain Feynman graph, 
by derivation with respect to one of the internal masses \cite{Kotikov} 
or one of the external invariants \cite{Remiddi} and with the repeated use 
of the integration by part identities, a system of independent first 
order partial differential equations is obtained in a number equal to 
the number of the MI (master differential equations). 

Enlarging the number of loops and legs grows the number of parameters, 
MI and equations, but does not change or spoil the method. 

To solve the system of equations (analytically or numerically) it is 
necessary to know the MI for a chosen value of the differential parameter. 
To achieve that analytically comes out to be the most laborious 
part of the method and often requires some external knowledge, 
like the assumption of regularity of the solution in the value. 
Moreover, if the chosen value is a zero for one of the coefficients of the 
derivative of the MI in the equations (as is always the case in 
analytical calculations), also the first derivative for that value is 
necessary to work out the numerical solution for different values of 
the parameter, but this usually comes out to be a simpler task 
(unless poles in the limit of the number of dimensions $n$ going to 
4 are present). 

\begin{figure}[htb]
\vspace{9pt}
\centering
{\scalebox{.7}[.7]{\includegraphics*[80,20][370,120]{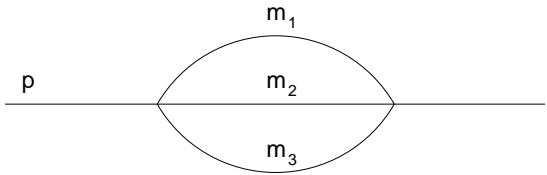}}}
\caption{The general massive 2-loop sunrise self-mass diagram.}
\label{fig:sunrise}
\end{figure}

To test the method we have chosen the simple, but not trivial, 2-loop 
sunrise graph with arbitrary masses \cite{CCLR1,CCR3}, shown in 
Fig.\ref{fig:sunrise}. 

This graph is one of the topologies of the 2-loop self-mass and has 
4 MI, the other topologies with 4 and 5 propagators have one more 
MI each \cite{Tarasov,CCLR2}. 

When the sunrise MI are expanded in $(n-4)$, the coefficients of 
the poles can be known analytically for arbitrary values of the 
external squared momentum $p^2$, while the finite parts satisfy 
the differential equations. 
From these equations the analytic expressions for their first order
expansion were completed at the {\em special} points 
\cite{CCLR1,CCR1,CCR2,CCR3}: 
$p^2=0$;  $p^2=\infty$;  $p^2=-(m_1+m_2+m_3)^2$, 
the threshold; $p^2=-(m_1+m_2-m_3)^2$, the pseudo-thresholds.

To obtain numerical results for arbitrary values of $p^2$, a 
4th-order Runge-Kutta formula is implemented in a FORTRAN code, 
with a solution advancing path starting from the {\em special} points, 
so that also the first term in the expansion is necessary.

The path followed starts usually from $p^2=0$ and moves in the lower 
half complex plane of $p^2$ to avoid proximity to the other {\em special} 
points, which can cause loss in precision. 
For values of $p^2$ very close to a {\em special} point, we start from 
the analytical expansion at that {\em special} point. 
Remarkable self-consistency checks are provided by choosing different 
paths or different starting points to calculate the same value.

The execution of the program is rather fast and precise: with an 
Intel Pentium III of 1 GHz we get values with 7 digits requiring 
times ranging from a fraction of a second to 10 seconds of CPU, 
and with 11 digits from few tens of seconds to one hour. 

If $\Delta=L/N$ is the length of one step, $L$ is the length of the 
whole path and $N$ the total number of steps, the 4th-order 
Runge-Kutta formula discards terms of order $\Delta^5$, so the 
whole error is $\epsilon_{RK} = N \Delta^5 = L^5/N^4$, and a proper 
choice of $L$ and $N$ allows the control of the precision. 

Indeed we estimate the relative error, as usual, by comparing a value 
obtained with $N$ steps with the one obtained with $N/10$ steps, 
$\epsilon_{RK} = [V(N)-V(N/10)]/V(N)$, to which we add a cumulative 
rounding error $\epsilon_{cre} = \sqrt{N} \times 10^{-15}$, 
due to our 15 digits double precision. 

Comparisons are done in \cite{CCR3} with some values present in the literature 
\cite{BBBS,Passarino} with excellent agreement.

\def\NP{{\sl Nucl. Phys.}} 
\def\PL{{\sl Phys. Lett.}} 
\def\PR{{\sl Phys. Rev.}} 
\def\PRL{{\sl Phys. Rev. Lett.}} 
\def\NC{{\sl Nuovo Cim.}}
\def\APP{{\sl Acta Phys. Pol.}}
\def\ZP{{\sl Z. Phys.}}
\def\MPL{{\sl Mod. Phys. Lett.}} 
\def\EPJ{{\sl Eur. Phys. J.}} 
\def\IJMP{{\sl Int. J. Mod. Phys.}}

\end{document}